%% file: main.tex
\documentclass[conference]{IEEEtran}
\IEEEoverridecommandlockouts
\usepackage{cite}
\usepackage{amsmath,amssymb,amsfonts}
\usepackage{algorithmic}
\usepackage{graphicx}
\usepackage{textcomp}
\usepackage{xcolor}
\usepackage{bbding}
\usepackage{url}
\usepackage{stfloats}
\usepackage{cite}
\usepackage{amsmath,amssymb,amsfonts}
\usepackage{algorithmic}
\usepackage{graphicx}
\usepackage{textcomp}
\usepackage{xcolor}
\usepackage{hyperref}
\usepackage{multirow}
\usepackage{threeparttable}
\usepackage{booktabs}
\usepackage{makecell}
\usepackage{colortbl}
\usepackage{tabularx}
\usepackage{bbding}
\usepackage{tabstackengine}
\usepackage{lipsum}
\usepackage{algorithm}
\usepackage{diagbox}

\def\BibTeX{{\rm B\kern-.05em{\sc i\kern-.025em b}\kern-.08em
    T\kern-.1667em\lower.7ex\hbox{E}\kern-.125emX}}

\begin{document}
\newcommand{\SolutionName}{DiffCkt}
\newcommand{\Code}{\url{https://github.com/CjLiu-NJU/DiffCkt}}

\DeclareRobustCommand*{\IEEEauthorrefmark}[1]{%
    \raisebox{0pt}[0pt][0pt]{\textsuperscript{\footnotesize\ensuremath{#1}}}}

\title{\SolutionName: A Diffusion Model-Based Hybrid Neural Network Framework
for Automatic Transistor-Level Generation of Analog Circuits
}

\author{
	\IEEEauthorblockN{
		Chengjie Liu\IEEEauthorrefmark{1,4}, 
		Jiajia Li\IEEEauthorrefmark{1}, 
		Yabing Feng\IEEEauthorrefmark{1}, 
		Wenhao Huang\IEEEauthorrefmark{3},\\
            Weiyu Chen\IEEEauthorrefmark{1,4},
		Yuan Du\IEEEauthorrefmark{1},
              Jun Yang\IEEEauthorrefmark{2,4},
        and Li Du\IEEEauthorrefmark{*1},
        }
	\IEEEauthorblockA{\IEEEauthorrefmark{1}School of Electronic Science and Engineering, Nanjing University, Nanjing, China}
 \IEEEauthorblockA{\IEEEauthorrefmark{2}School of Integrated Circuits, Southeast University, Nanjing, China}
 \IEEEauthorblockA{\IEEEauthorrefmark{3}School of Integrated Circuits, Nanjing University, Nanjing, China}
    \IEEEauthorblockA{\IEEEauthorrefmark{4}National Center of Technology Innovation for EDA, Nanjing, China}
     \IEEEauthorblockA{\IEEEauthorrefmark{*}Corresponding Author: Li Du \quad Email: ldu@nju.edu.cn}}

\maketitle
\vspace{-50pt}

\begin{abstract}

Analog circuit design consists of the pre-layout and layout phases. Among them, the pre-layout phase directly decides the final circuit performance, but heavily depends on experienced engineers to do manual design according to specific application scenarios.
To overcome these challenges and automate the analog circuit pre-layout design phase, we introduce {\SolutionName}: a diffusion model-based hybrid neural network framework for the automatic transistor-level generation of analog circuits, which can directly generate corresponding circuit structures and device parameters tailored to specific performance requirements.
To more accurately quantify the efficiency of circuits generated by {\SolutionName}, we introduce the Circuit Generation Efficiency Index (CGEI), which is determined by both the figure of merit (FOM) of a single generated circuit and the time consumed. Compared with relative research, {\SolutionName} has improved CGEI by a factor of $2.21 \sim 8365 \times$ , reaching a state-of-the-art (SOTA) level.
In conclusion, this work shows that the diffusion model has the remarkable ability to learn and generate analog circuit structures and device parameters, providing a revolutionary method for automating the pre-layout design of analog circuits. The circuit dataset will be open source, its preview version is available at {\Code}.
\end{abstract}
\vspace{-5pt}
\section{Introduction}
\input{Intro/Intro.tex}
\section{Preliminaries}
\input{BackGround/BackGround.tex}
\section{Amplifier Data Construction}
\input{Amplifier_Data_Construction/AmplifierDataConstruction.tex}
\section{\SolutionName}
\input{DiffCkt/DiffCkt.tex}
\section{Experiment}
\input{Experiment/Experiment.tex}
\section{Conclusion and Discussion}
\input{Conclusion/conclusion.tex}
\bibliography{main.bbl}
\bibliographystyle{IEEEtran}

\end{document}

%% file: Intro/Intro.tex
Analog circuits are an essential component of integrated circuits (ICs). Nevertheless, their design process, encompassing the pre-layout phase aspects including topology selection and device parameter sizing, along with layout placement and routing, is less automated when contrasted with digital circuits. Among them, compared with the layout phase that already has a certain degree of automation\cite{chang2018bag2,xu2019magical}, the pre-layout phase of analog circuits still requires expert experience and is basically implemented purely manually. Analog circuit engineers need to select the most suitable topology structure based on the metric requirements, and this topology structure often determines the final range of circuit metrics.

At the pre-layout design phase, numerous studies have explored the use of optimization algorithms like Bayesian optimization\cite{pmlr-v80-lyu18a} or reinforcement-learning-based methods \cite{wang2020gcn}
to automate the parameter sizing of analog circuits, ultimately achieving device parameters that satisfy performance criteria. 
However, these approaches are limited to device parameter optimization for fixed circuit structures and fail to achieve the generation of circuit structures based on specific performance requirements. Additionally, due to the necessity of iterative simulations, these methods incur substantial time costs.

\renewcommand{\arraystretch}{1} 
\begin{table}[!t]
\centering
\caption{Comparison of \textbf{{\SolutionName}} with similar Analog EDA works}
\label{tab:comparison}
\begin{tabular}{cccc}
\toprule
\textbf{Approach} & \textbf{\makecell{No Need for \\ Structural Input?}} & \textbf{\makecell{No Need for \\ Simulation \\ Iteration?}} & \textbf{\makecell{Transistor-Level \\ Generation?}} \\
\midrule
MACE\cite{pmlr-v80-lyu18a} & \textcolor{red}{\text{\sffamily \XSolidBrush}} & \textcolor{red}{\text{\sffamily \XSolidBrush}} & — \\
GCN-RL\cite{wang2020gcn} & \textcolor{red}{\text{\sffamily \XSolidBrush}} & \textcolor{red}{\text{\sffamily \XSolidBrush}} & — \\
AmpAgent\cite{liu2024ampagent} & \textcolor{red}{\text{\sffamily \XSolidBrush}} & \textcolor{red}{\text{\sffamily \XSolidBrush}} & — \\
CktGNN\cite{dong2024cktgnncircuitgraphneural} & \textcolor{green!50!black}{\text{\sffamily \checkmark}} & \textcolor{red}{\text{\sffamily \XSolidBrush}} & \textcolor{red}{\text{\sffamily \XSolidBrush}} \\
Atom\cite{shen2024atom} & \textcolor{green!50!black}{\text{\sffamily \checkmark}} & \textcolor{red}{\text{\sffamily \XSolidBrush}} & \textcolor{green!50!black}{\text{\sffamily \checkmark}} \\
\rowcolor{green!20}\textbf{DiffCkt} & \textcolor{green!50!black}{\text{\sffamily \checkmark}} & \textcolor{green!50!black}{\text{\sffamily \checkmark}} & \textcolor{green!50!black}{\text{\sffamily \checkmark}} \\
\bottomrule
\end{tabular}
\end{table}


\cite{dong2024cktgnncircuitgraphneural, shen2024atom, poddar2024data} also achieved the automatic generation of analog structures based on specific performance requirements. However, \cite{dong2024cktgnncircuitgraphneural} remained at the stage of ideal modules. \cite{shen2024atom, poddar2024data} relied on a fixed post-processing scheme, which means specific transistor structures were used to replace each sub-module, resulting in a lack of scalability and diversity.
Moreover, \cite{dong2024cktgnncircuitgraphneural, shen2024atom, poddar2024data} still rely on other optimization algorithms\cite{pmlr-v80-lyu18a} to achieve parameter tuning, which causes the problem of long processing times.

Recent research investigated the application of large language models (LLMs) for analog circuits auto-generation \cite{liu2024ladac,lai2024analogcoder,yin2024ado,chang2024lamagic,liu2024ampagent,gao2025analoggenie} and demonstrated high efficiency. Nonetheless, because analog circuits are inherently graph-based \cite{9218757}, as opposed to the sequential data that LLMs are designed to handle, LLMs must rely on circuit netlists and other natural language forms for analog circuit representation. This approach often includes extraneous information, leading to reduced representational efficiency. Furthermore, LLMs have very large model parameters, which also incur significant overhead during training.  

In summary, the existing analog circuit automation techniques face issues such as a lack of flexibility in generated structures and long generation times.

Actually, the problem of generating a circuit structure and device parameters from given specifications inherently belongs to the category of inverse problems, for which the diffusion models stand as a robust approach.
Diffusion models have demonstrated exceptional performance in the field of Text to Image (T2I) generation\cite{sohl2015deep,ho2020denoisingdiffusionprobabilisticmodels,ho2022imagenvideohighdefinition,rombach2022highresolutionimagesynthesislatent,guo2023animatediff}, leading to the development of mature commercial models such as DALL·E 3\footnote{\url{https://openai.com/index/dall-e-3/}}. 
Moreover, in graph  generation problems, where the representation methods are equivalent to those utilized for analog circuits, as exemplified by applications such as protein generation and molecular docking, the Diffusion model \cite{chamberlain2021grand,liu2023generativediffusionmodelsgraphs,vignac2023digressdiscretedenoisingdiffusion} showcases exceptional performance and outperforms traditional GNN models. This leads us to believe that diffusion models can also learn the connections between circuit devices and their parameters, thereby enabling the generation of corresponding circuit structures and device parameters based on specified performance requirements.

Based on this foundation, we propose {\SolutionName}: a diffusion model-based hybrid neural network framework for automatic transistor-level generation of analog circuits. To the best of our knowledge, compared with similar works\cite{eid2025using,de2025comprehensive}, {\SolutionName} is the first work to adopt the diffusion model, enabling the automatic design of analog circuits at the transistor level, covering from structure generation to sizing.
{\SolutionName} takes the circuit performance requirements as input and generates the circuit structure and its devices' parameters.
To achieve this, {\SolutionName} is composed of 3 neural networks:
A Multilayer Perceptron (MLP) network for the circuit's components number prediction;
A discrete denoising diffusion network for the circuit's components' type and connection prediction;
A continuous denoising diffusion network for components' sizing.
By connecting these three networks in series, {\SolutionName} enables the generation of circuit structures and device parameters based on the performance specifications.

To achieve an efficient representation of analog circuits, we proposed a novel graph format specifically designed to describe device attributes and their connections. 
{\SolutionName} is trained on this formatted data to develop a foundational capability for generating circuit structures. 
The dataset is obtained by exhaustively sampling and simulating various operational amplifier structures and parameters on the TSMC 65nm CMOS process. Ultimately, we acquired over 400k sets of amplifier structure-performance metric pairs.

To more accurately quantify the efficiency of circuits generated by {\SolutionName}, we introduce the Circuit Generation Efficiency Index (CGEI), which is determined by both the figure of merit (FOM) of a single generated circuit and the time consumed. Compared with relative research, {\SolutionName} has improved CGEI by a factor of $2.21 \sim 8365 \times$, reaching a state-of-the-art (SOTA) level.

In summary, we are the first to introduce the use of diffusion models for the automatic generation of analog circuits. This work highlights the diffusion model's ability to learn analog circuit structures and device parameters, offering a novel approach to advancing analog circuit automation. This paper is organized as follows: In Section 2, we will formulate the problem and introduce the foundational concepts. In Section 3, we will discuss the data sampling methods and provide detailed descriptions. Section 4 will detail the {\SolutionName} framework. Section 5 will present the experiments. Finally, Section 6 will conclude the paper and outline future work plans. The circuit dataset will be open source, its preview version is available at {\Code}.

%% file: BackGround/BackGround.tex
In this section, we will first introduce why choosing amplifiers as outer research objects and our method of converting circuits into graph representations. Next, we will explain how diffusion models can be utilized to learn representations of both the structure and device parameters of analog circuits. Finally, we will present our approach to modeling the problem of learning analog circuit representations using diffusion models.
\begin{figure*}[!t]
\setlength{\abovecaptionskip}{0cm}
\setlength{\belowcaptionskip}{-10cm}
  \centering
  \includegraphics[width=0.9\linewidth]{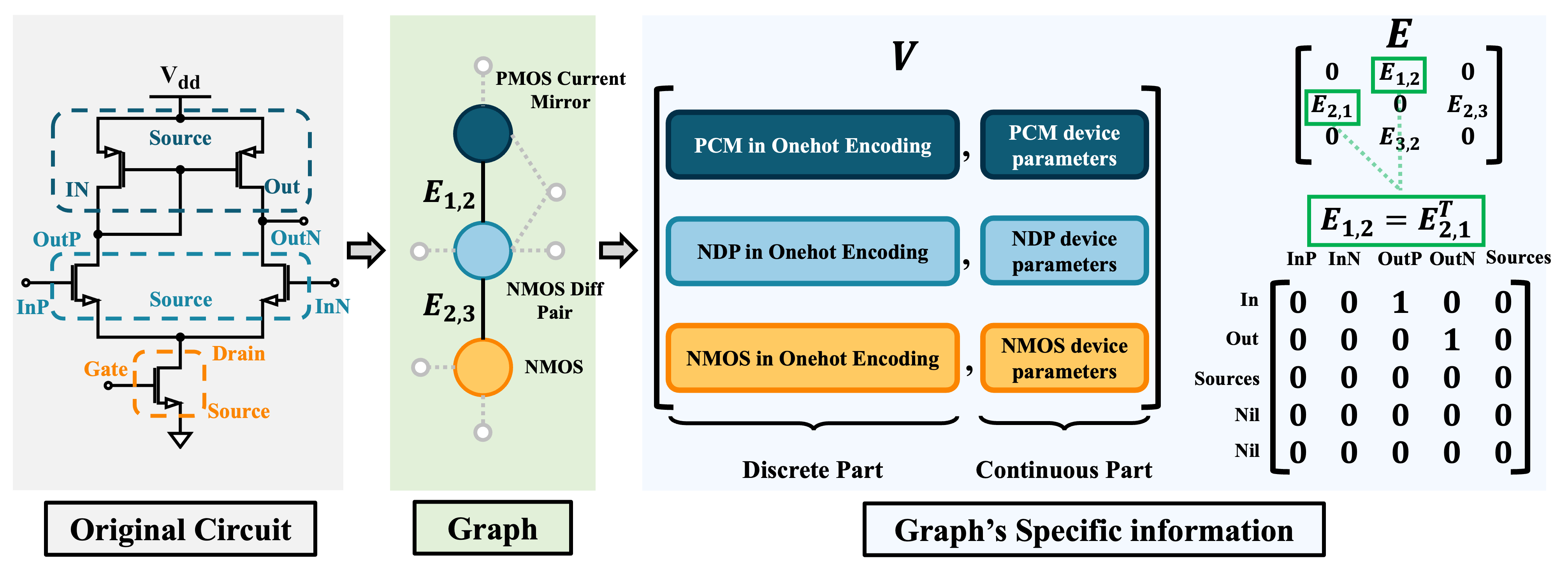}
  \caption{An example of transferring an amplifier into its corresponding graph representation.}
  \label{fig:graph_representation}
  \vspace{-15pt}
\end{figure*}

\subsection{Amplifiers}
Amplifiers, being the most fundamental circuit type, are extensively employed in various types of analog circuits and mixed-signal modules. The amplifiers are incorporated into the vast majority of circuits, such as sensors, filters, and analog-to-digital converters (ADCs). However, amplifiers are a typical type of circuit constrained by the octagon rule\cite{allen2011cmos}, which implies that when high performance is demanded, designers typically devote considerable time to selecting an optimal amplifier structure and sizing the components to ultimately fulfill the performance criteria. 

Therefore, our aim is to develop a neural network framework that can comprehensively learn the correlations between amplifier performance metrics and various amplifier structures and their device parameters. This would allow for the input of specific performance metrics and the output of appropriate amplifier structures and device parameters.

\subsection{Graph Representation}
Many previous works have explored using graphs to represent circuits\cite{9218757,dong2024cktgnncircuitgraphneural}, viewing the entire circuit as a graph \( \boldsymbol{G} \) with \( n \) nodes, where devices are represented as nodes \( \boldsymbol{V} \in \mathbb{R}^{n{\times}a} \), with \( a \) being the dimensionality of the nodes' one-hot encoding, and the connections between devices as edges \( \boldsymbol{E} \in \mathbb{R}^{n{\times}n} \) connecting the nodes. Although circuits can indeed be represented in graph form, only constructing the edges \( \boldsymbol{E} \) presents challenges because the connections between different devices cannot be adequately captured using a simple binary representation of connectivity. This limitation arises because the connections to different ports of devices exhibit distinct electrical characteristics, leading to shortcomings in the current methods for effectively representing circuits as graphs.

To achieve a more comprehensive graph representation of analog circuits, we have optimized the method of representing edges. The attributes of each edge will be represented by a matrix \( \boldsymbol{\xi} \in \mathbb{R}^{ k \times k}\), where $k$ is the maximum port number of all types of components. When \( \boldsymbol{\xi}_{i,j} \) is 1, it indicates that in the matrix \( \boldsymbol{\xi} \), the \( i \)-th port of the first device is connected to the \( j \)-th port of the second device. Conversely, if \( \boldsymbol{\xi}_{i,j}\) is 0, it indicates that these two ports are not connected. Additionally, to incorporate more prior knowledge of analog circuits, this method treats essential analog circuit building blocks, such as differential pairs and current mirrors, as single nodes, since they often share the same parameters.

An example is shown in Fig.\ref{fig:graph_representation}, where we transfer an NMOS-input 5-transistor OTA into its corresponding graph representation. First, the components are divided into a \texttt{PMOS Current Mirror}, an \texttt{NMOS Differential Pair}, and a single \texttt{NMOS}.

Next, each component will be one-hot encoded to form the discrete part of the node matrix \( \boldsymbol{V} \). The device-specific parameters, such as the channel width and channel length of a MOS transistor, will form the continuous part of the node matrix \( \boldsymbol{V} \). The edge attributes between adjacent nodes will be recorded, with matrices corresponding to symmetrical edges being transposed of each other. For unconnected nodes, the edge attributes will be zero matrices. For example, consider the connection between the \texttt{NMOS differential pair} and the \texttt{PMOS current mirror}. The output on the left side of the differential pair, which we define as the \texttt{OutP} port, connects to the \texttt{IN} port of the \texttt{PMOS current mirror}. They separately correspond to the first and third ports of their respective devices, so \( \boldsymbol{\xi}_{1,3} \) in \( \boldsymbol{E}_{1,2} \) is 1. Similarly, the \texttt{OutN} port of the differential pair connects to the OUT port of the current mirror. both corresponding to the second port of their respective devices, making \( \boldsymbol{\xi}_{2,4} \) in \( \boldsymbol{E}_{1,2} \) equal to 1. No other ports are connected between these two devices, so all other elements in \( \boldsymbol{E}_{1,2} \) are 0.

With this approach, we can effectively represent the types of devices, their parameters, and their interconnections.

\subsection{Problem Formulation}
The goal of our model is to predict a suitable amplifier structure and the parameters of each device within it based on input specifications. With this consideration, we have defined three problem scenarios:

$Problem 1$: Given the specification requirements, predict the number of devices \( N \) in the circuit, and generate a graph with \( N \) nodes that represent the devices in the circuit, each with random types and connections.

$Problem 2$: Given the specification requirements, predict the device type and connection in the amplifier. This involves predicting node types and edges in the discrete domain, ultimately providing an amplifier structure with randomly assigned device parameters as a graph.

$Problem 3$: Given the specification requirements and the graph corresponding to the amplifier structure, predict the values for each device to meet the amplifier's performance requirements.

Accordingly, our model, {\SolutionName}, is designed to sequentially address these three problems, ultimately providing an amplifier structure and its corresponding device parameters that satisfy the specified performance requirements.

%% file: Amplifier_Data_Construction/AmplifierDataConstruction.tex
In this section, we will explain how we constructed the amplifier dataset and describe the distribution of the final dataset.
\subsection{Dataset construction}
\label{Sec:dataset construction}
We performed random sampling across different amplifier structures and device parameters to ensure diversity in the dataset regarding circuit structures and parameters. Under five different multistage amplifier topologies, by restructuring with eight different single-stage amplifiers, we finally generated 28 different amplifier structures. The parameters of each component in the amplifier were randomly sampled within their respective fixed ranges. 

After sampling the structure and parameters of an amplifier, we simulate it using the Spectre$^\circledR$ simulator to obtain its specific parameters. The simulation will generate the following metrics for the amplifier: 
\texttt{power consumption ($P_{diss}$)}, 
\texttt{DC gain ($Gain_{DC}$)}, 
\texttt{gain-bandwidth product ($GBW$)}, 
\texttt{phase margin ($PM$)}, 
\texttt{positive slew rate ($SR_P$)}, 
\texttt{negative slew rate ($SR_N$)}, 
\texttt{output voltage swing low ($VOL$)}, 
\texttt{output voltage swing high ($VOH$)}, 
\texttt{common-mode rejection ratio ($CMRR$)}, 
\texttt{power supply rejection ratio ($PSRR$)}, 
\texttt{input equivalent noise at 1 kHz ($Noise_{@1kHz}$)}, 
\texttt{input equivalent noise at 1 GHz($Noise_{@1GHz}$)}, 
and \texttt{load cap ($C_L$)}. 
In total, there are 13 performance metric requirements. When applying {\SolutionName}, these performance metric requirements are also used as inputs.
\subsection{Dataset Distribution}

\begin{figure}[!t]
  \centering
  \includegraphics[width=0.8\linewidth]{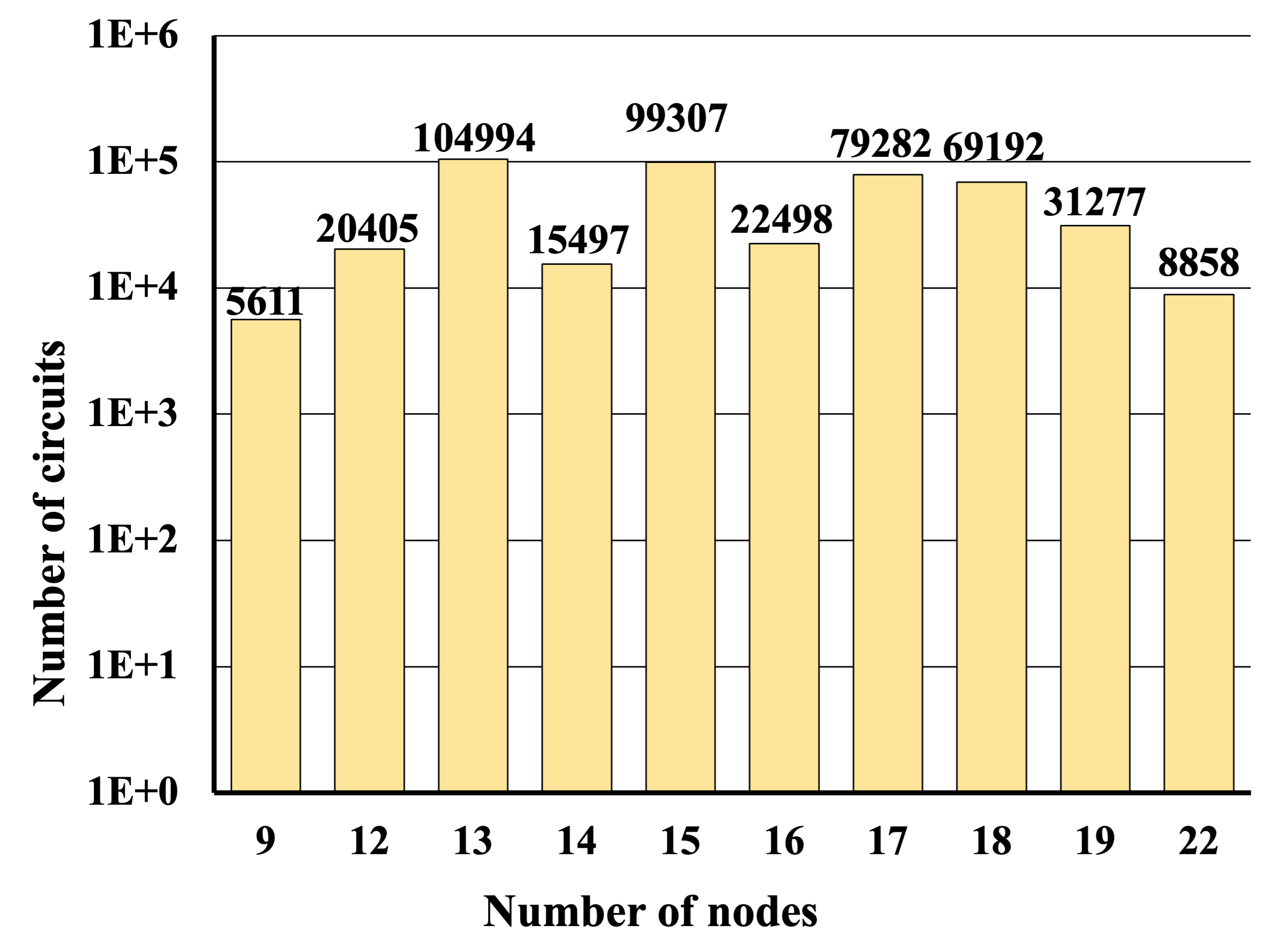}
  \caption{Data distribution in our datasets.}
  \label{fig:Node number in training dataset}
\end{figure}

We sampled approximately 60k data points for single-stage amplifiers and around 90k data points for each type of multi-stage amplifier. All these data were constructed based on TSMC65. Ultimately, we constructed a dataset of over 400k data points, with their distribution according to the number of nodes shown in Fig.\ref{fig:Node number in training dataset}. The circuit dataset will be open source, its preview version is available at {\Code}.

%% file: DiffCkt/DiffCkt.tex
\begin{figure*}[!t]
  \centering
  \includegraphics[width=0.9\linewidth]{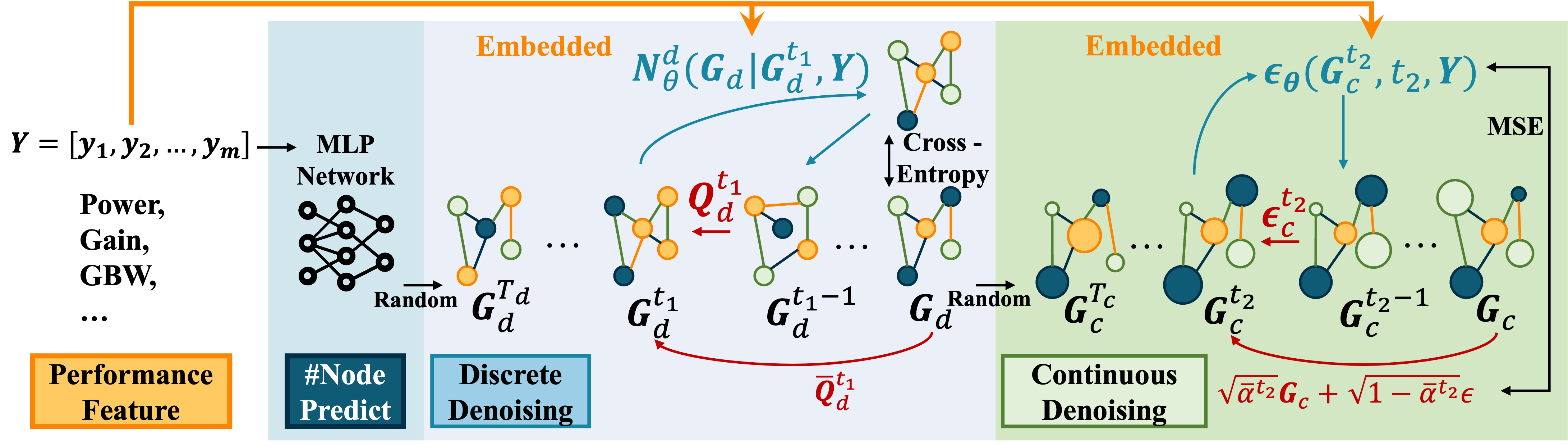}
  \caption{The overview of DiffCkt.}
  \label{fig:DiffCkt}
  \vspace{-15pt}
\end{figure*}
In this section, we will introduce each module in the {\SolutionName} workflow. We will also describe the diffusion and denoising processes involved. 

The matrix \( \boldsymbol{Y} \in \mathbb{R}^{m} \) represents the requirement matrix, where \( m \) denotes the length of the requirements, documenting the detailed specification demands. \( \boldsymbol{G}_d \) denotes the discrete graph. 
Within this graph, the node matrix \( \boldsymbol{V}_d \in \mathbb{R}^{n \times a} \) is used, where each node \( \boldsymbol{V}_{d,i} \in \mathbb{R}^{a}\) is one-hot encoded, with \( a \) being the dimensionality of the nodes' one-hot encoding and \( n \) being the number of nodes. 
Additionally, there is an adjacency matrix \( \boldsymbol{E}_d \in \mathbb{R}^{n \times n} \), where \( \boldsymbol{E}_{d_{i,j}} \in \mathbb{R}^{k \times k }\).

Similarly, \( \boldsymbol{G}_c \) is a continuous graph, and the node matrix \( \boldsymbol{V}_c \) is a matrix in \( \mathbb{R}^{n \times (a+b)} \), where \( b \) is the dimension of the continuous parameters. Each node \( \boldsymbol{V}_{c,i} \) is a vector in \( \mathbb{R}^{(a+b)} \). The dimensions of the adjacency matrix remain unchanged.

\subsection{{\SolutionName}'s overview}
The {\SolutionName}'s overview is shown in Fig.\ref{fig:DiffCkt}. {\SolutionName} is composed of 3 networks: a Multilayer Perceptron (MLP)-based network for predicting the number of devices, a discrete-graph denoising diffusion model for the graph's nodes classification and edge prediction, and a continuous-graph denoising diffusion model for devices' parameters prediction. 

When a specific set of specification requirements is input, represented by a vector \( \boldsymbol{Y} \), the first network predicts the number of nodes \( n \). Subsequently, a random discrete matrix \( \boldsymbol{G}_d \) with \( n \) nodes is generated, where each node has a random type and random interconnections. The discrete denoising diffusion network then progressively denoises this random discrete matrix while taking \( \boldsymbol{Y} \) into account, ultimately predicting a suitable circuit structure. This involves accurately predicting the node types and their interconnections in \( \boldsymbol{G}_d \). Following this step, a random continuous feature matrix \( \boldsymbol{G}_c \) is generated, essentially assigning random parameters to each device. The continuous denoising diffusion network, using \( \boldsymbol{Y} \), progressively denoises this random continuous matrix to predict the device parameters that meet the specifications, effectively determining the node parameters in \( \boldsymbol{G}_c \).

\subsection{Diffusion process and denoising iterations}


\begin{algorithm}[!t]
\caption{Training Process on Continuous Graph}
\label{alg:Continuous_Training}
\begin{algorithmic}[1]
\STATE {\bfseries Input:} Continuous Graph $\boldsymbol{G}_c$ and its Nodes $\boldsymbol{V}_c$, Requirement Matrix $\boldsymbol{Y}$, Number of epochs $N_{\text{epoch}}$, Total timesteps $T$
\FOR{$i = 1$ {\bfseries to} $N_{\text{epoch}}$}
    \STATE Randomly sample $t \sim \text{Uniform}(\{1, 2, \ldots, T\})$
    \STATE $\boldsymbol{V}_c^t \leftarrow  \sqrt{\overline{\alpha}^t} \boldsymbol{V}_c + \sqrt{1-\overline{\alpha}^t} \boldsymbol{\epsilon}$
    \STATE $\hat{\boldsymbol{\epsilon}} \leftarrow  \text{DiffCkt}(\boldsymbol{G}_c^t, Y, t)$
    \STATE Compute loss: $loss \leftarrow  MSE(\hat{\boldsymbol{\epsilon}},\boldsymbol{\epsilon})$
    \STATE Update model parameters
\ENDFOR
\end{algorithmic}
\end{algorithm}

For image diffusion, Gaussian noise $\boldsymbol{\varepsilon}$ is typically added to progressively obscure the original image information \cite{sohl2015deep, ho2020denoisingdiffusionprobabilisticmodels}. A denoising network is then trained to predict the noise distribution added at each step. We employ a similar approach for continuous graphs, where noise $\boldsymbol{\varepsilon}_c^t$ is added to the continuous parameters of the nodes. A continuous graph denoising network is trained to predict the noise added to these continuous node parameters. 

\begin{algorithm}[!t]
\caption{Training Process on Discrete Graph}
\label{alg:Discrete_Training}
\begin{algorithmic}[1]
\STATE {\bfseries Input:} Discrete graph $\boldsymbol{G}_d$, and its Node vector $\boldsymbol{V}$ and Adjacency matrix $\boldsymbol{E}$,  Requirement Matrix $\boldsymbol{Y}$, Number of epochs $N_{\text{epoch}}$,  Total timesteps $T$
\FOR{$i = 1$ {\bfseries to} $N_{\text{epoch}}$}
    \STATE Randomly sample $t \sim \text{Uniform}(\{1, 2, \ldots, T\})$
    \STATE $\overline{\boldsymbol{Q}}^t \leftarrow  \overline{\alpha}^t \boldsymbol{I} + (1 - \overline{\alpha}^t) \boldsymbol{1}_d\boldsymbol{1}_d^T/d$ 
    \STATE $\boldsymbol{V}^{t} \leftarrow  \boldsymbol{V} \cdot \overline{\boldsymbol{Q}}_{d,V}^t$
    \FOR{each $\boldsymbol{\xi}$ in $\boldsymbol{E}$}
        \STATE $\boldsymbol{\xi}^{t} \leftarrow  \boldsymbol{\xi} \cdot \overline{\boldsymbol{Q}}_{d,\xi}^{t}$
    \ENDFOR
    \STATE $\boldsymbol{G}^t \leftarrow  (\boldsymbol{V}^t,\boldsymbol{E}^t)$
    \STATE $(\hat{p}_{\boldsymbol{V}},\hat{p}_{\boldsymbol{E}}) \leftarrow  \text{DiffCkt}(\boldsymbol{G}^t, Y, t)$ 
    \STATE Compute loss:
    $loss \leftarrow  \text{CrossEntropy}(\hat{p}_{\boldsymbol{V}}, \boldsymbol{V}) + \sum_{\hat{p}_{\boldsymbol{\xi}} \in \hat{p}_{\boldsymbol{E}}} \text{BinaryCrossEntropy}(\hat{p}_{\boldsymbol{\xi}}, \boldsymbol{\xi})$
    \STATE Update model parameters
\ENDFOR
\end{algorithmic}
\end{algorithm}

In the diffusion process applied to the continuous parameters of a graph's node matrix, the progressive addition of noise at each timestep \( t \) is expressed as:

\begin{align}
\boldsymbol{V}_c^{t} &= \sqrt{\alpha^t} \boldsymbol{V}_c^{t-1} + \sqrt{1-\alpha^t} \boldsymbol{\epsilon}^t
\notag
\\&= \sqrt{\overline{\alpha}^t} \boldsymbol{V}_c + \sqrt{1-\overline{\alpha}^t} \boldsymbol{\epsilon}
\label{eq:diffusion_continuous}
\end{align}

In this equation, \(\boldsymbol{V}_c^{(t)}\) represents the noisy node matrix at timestep \( t \), where $\boldsymbol{V}_c^{(t-1)}$ is the previous noisy node matrix, and \(\boldsymbol{\epsilon}^{(t)} \sim \mathcal{N}(0, \boldsymbol{I})\) is the Gaussian noise introduced at this timestep. The parameter \(\alpha^t\) determines the proportion of the original signal retained versus the noise incorporated. This formulation effectively diffuses the original data into noise over the specified timesteps, and this process is crucial in training denoising models to reverse the diffusion and recover the original node parameters.

For discrete graphs, due to the discrete nature of node types and connections, the noise addition process involves modifying the probability distribution of each node and edge by the transition probabilities matrix $\boldsymbol{Q}_{d,V}^t$ and $\boldsymbol{Q}_{d,\xi}^t$ \cite{vignac2023digressdiscretedenoisingdiffusion}. This allows for sampling a noisy version of the discrete graph. The $\boldsymbol{Q}^t$ is defined as
\begin{equation}
\boldsymbol{Q}^t = \alpha^t \boldsymbol{I} + (1 - \alpha^t) \frac{1}{d} \boldsymbol{1}_d\boldsymbol{1}_d^T
\end{equation}
where \(\boldsymbol{I}\) is the identity matrix, \( d \) represents either the number of one-hot encoded device types or the number of \(\xi\)'s ports number in the adjacency matrix $\boldsymbol{E}$, and \(\boldsymbol{1}\) is a vector of ones.
And we have 
\begin{equation}
\overline{\boldsymbol{Q}}^t = \overline{\alpha}^t \boldsymbol{I} + (1 - \overline{\alpha}^t) \frac{1}{d} \boldsymbol{1}_d\boldsymbol{1}_d^T
\end{equation}
For the node vector $\boldsymbol{V}$, we apply $\boldsymbol{V}^{t}=\boldsymbol{V}^{t-1} \cdot \boldsymbol{Q}_{d,V}^{t} = \boldsymbol{V}^0 \cdot \overline{\boldsymbol{Q}}_{d,V}^t$.
For every $\boldsymbol{\xi}$ in $\boldsymbol{E}$, we apply $\boldsymbol{\xi}^{t}=\boldsymbol{\xi}^{t-1} \cdot \boldsymbol{Q}_{d,\xi}^{t} = \boldsymbol{\xi}^0 \cdot \overline{\boldsymbol{Q}}_{d,\xi}^{t}$.
Finally, sampling will be conducted based on the values of each node and edge to obtain the noisy discrete graph, \( \boldsymbol{G}^t \).
The specific training procedures for continuous and discrete denoising are separately shown in the Algorithm \ref{alg:Continuous_Training} and Algorithm \ref{alg:Discrete_Training}.

\begin{algorithm}[!t]
\caption{Sampling for Continuous Graph Diffusion}
\label{alg:Sampling_Continuous}
\begin{algorithmic}[1]
\STATE {\bfseries Input:} Fix-structure but with noisy continuous node attribute Graph $\boldsymbol{G}_c$ and its continuous node attribute $\boldsymbol{V}_c^T$, Requirement Matrix $\boldsymbol{Y}$, Total timesteps $T$
\STATE Initialize model parameters and data structures
\FOR{$t = T$ {\bfseries downto} $1$}
    \STATE Predict noise: $\hat{\boldsymbol{\epsilon}}^t \leftarrow \text{DiffCkt}(\boldsymbol{G}_c^t, \boldsymbol{Y}, t)$
    \STATE $\boldsymbol{V}_c^{t-1} \leftarrow (\boldsymbol{V}_c^t - \frac{1-\alpha^t}{\sqrt{1-\overline{\alpha}^t}}\hat{\boldsymbol{\epsilon}}^t) / {\sqrt{\alpha^t}}$
    \IF{$t > 1$}
        \STATE Store or further process $\boldsymbol{V}_c^{t-1}$
    \ENDIF
\ENDFOR
\STATE {\bfseries Output:} Denoised continuous node attribute $\boldsymbol{V}_c^0$
\end{algorithmic}
\end{algorithm}


For continuous graph diffusion within the {\SolutionName} framework, the sampling process is designed to iteratively refine the noisy continuous node attributes of a fixed-structure graph. The process begins with a graph \(\boldsymbol{G}_c\) characterized by its noisy node attributes \(\boldsymbol{V}_c^T\) and a requirement matrix \(\boldsymbol{Y}\). Over a total of \(T\) timesteps, the algorithm systematically reduces noise, thereby reconstructing the original node attributes. At each timestep \(t\), the DiffCkt model predicts the noise component \(\hat{\boldsymbol{\epsilon}}^t\) by leveraging the current graph state \(\boldsymbol{G}_c^t\), the requirement matrix \(\boldsymbol{Y}\), and the timestep \(t\). This predicted noise is then used to update the node attributes according to the equation \(\boldsymbol{V}_c^{t-1} \leftarrow (\boldsymbol{V}_c^t - \frac{1-\alpha^t}{\sqrt{1-\overline{\alpha}^t}}\hat{\boldsymbol{\epsilon}}^t) / {\sqrt{\alpha^t}}\), where \(\alpha^t\) is a predefined parameter that controls the denoising strength at each step. This iterative process continues until \(t = 1\), at which point the algorithm outputs the denoised continuous node attributes \(\boldsymbol{V}_c^0\). This approach effectively reconstructs the original node attributes by progressively removing noise, thereby enhancing the fidelity of the graph's representation.

For discrete graph diffusion, the process starts with a noisy discrete graph \(\boldsymbol{G}_d^T\) comprising node attributes \(\boldsymbol{V}_d^T\) and edge attributes \(\boldsymbol{E}_d^T\), along with the requirement matrix \(\boldsymbol{Y}\). The process iterates over \(T\) timesteps, from \(T\) down to 1. At each timestep \(t\), {\SolutionName} estimates probabilities \(\hat{p}_V\) and \(\hat{p}_E\) for nodes and edges, respectively, based on the current graph state \(\boldsymbol{G}_d^t\) and \(\boldsymbol{Y}\). For each node \(i\), the probability of its attribute at \(t-1\) is updated using \(p_\theta(\boldsymbol{V}^{t-1}_i \mid \boldsymbol{G}^t) \leftarrow \sum _x q(x_i^{t-1}\mid x_i = x,x_i^t) \hat{p}_i^V(x)\). For each edge \(i, j\) and each element \(\xi^{t}_{u, v}\), the probability of its attribute at \(t-1\) is updated using \(p_\theta(\xi^{t-1}_{u, v} \mid \boldsymbol{G}^t) \leftarrow \sum_{e} q(\xi_{u, v}^{t-1} \mid \xi_{u, v} = e, \xi_{u, v}^{t}) \hat{p}_{ij}^{E}(e)\), where \(e \in \{0,1\}\). The process outputs an approximate original graph \(\boldsymbol{G}_d^0\). These processes effectively reconstruct both continuous and discrete graph structures from noisy inputs using the {\SolutionName}.

The sampling procedure in discrete and continuous noise-denoising diffusion models are detailed in Algorithm \ref{alg:Sampling_Continuous} and Algorithm \ref{alg:Sampling_Discrete}, respectively.

\begin{algorithm}[!t]
\caption{Sampling for Discrete Graph Diffusion}
\label{alg:Sampling_Discrete}
\begin{algorithmic}[1]
\STATE {\bfseries Input:} Noisy Discrete Graph $\boldsymbol{G}^T_d = (\boldsymbol{V}_d^T,\boldsymbol{E}_d^T)$, Requirement Matrix $\boldsymbol{Y}$, Total timesteps $T$
\FOR{$t = T$ {\bfseries downto} $1$}
    
    \STATE $\hat{p}_V, \hat{p}_E \leftarrow  \text{DiffCkt}(\boldsymbol{G}_d^t, Y, t)$
    
    \STATE $p_\theta(\boldsymbol{V}^{t-1}_i \mid \boldsymbol{G}^t) \leftarrow \sum _x q(x_i^{t-1}\mid x_i = x,x_i^t) \hat{p}_i^V(x) $\\
    $i \in 1,\dots,n$
    
    \FOR{each edge $i, j = 1, \ldots, n$}
    \FOR{each element $\xi^{t}_{u, v} \in \boldsymbol{E}_{i,j}^t$}
        \STATE $p_\theta(\xi^{t-1}_{u, v} \mid \boldsymbol{G}^t) \leftarrow 
        \sum_{e} q(\xi_{u, v}^{t-1} \mid \xi_{u, v} = e, \xi_{u, v}^{t}) \hat{p}_{ij}^{E}(e)  
        $ \\
        $e \in \{0,1\}$
    \ENDFOR
\ENDFOR
\ENDFOR
\STATE {\bfseries Output:} Approximate original graph $\boldsymbol{G}_d^0$
\end{algorithmic}
\end{algorithm}

%% file: Experiment/Experiment.tex
\input{Experiment/normal_tab.tex}
\input{Experiment/normal_table.tex}
In this section, we will first introduce the experimental setup. Secondly, we will adopt the DDIM\cite{song2022denoisingdiffusionimplicitmodels} approach to observe various performances of {\SolutionName} under different sampling interval steps. Furthermore, we will assess the quality of the generated circuits under different requirements and their performance in relation to these requirements. Finally, we will evaluate the comparison between {\SolutionName} and relevant reproducible works.

\subsection{Experiment Setup}
During the training phase, the node number prediction network in {\SolutionName} was trained on a single NVIDIA A800 GPU, while both the discrete denoising network and the continuous denoising network were trained on three NVIDIA A800 GPUs each. The models were trained with a batch size of 8192 over 1000 iterations. The total diffusion step is set as 500.

Additionally, to accelerate inference speed, we employed the DDIM\cite{song2022denoisingdiffusionimplicitmodels} method and set the number of interval steps to 1, 5, 10, and 20. We will demonstrate the impact of setting different interval steps on the impact on the generated circuits' performance and time consumption.

The performance metrics' normalization parameters are shown in Table.\ref{tab:Normal parameters_II}. The normalized sampling range is presented in Table.\ref{tab:normalized}. The sampled parameters will be used as inputs for {\SolutionName}. The sampling space is divided into four parts. Three of them conform to the distribution of the training data indicators, namely Low, Medium, and High. There is also an \textbf{External} part that slightly exceeds the data distribution of the training set. This part is used to observe how {\SolutionName} performs when faced with circuit performance ranges beyond those of the training set. 

\begin{algorithm}[!t]
\caption{Performance Evaluation}
\label{alg:Dynamic Loss}
\begin{algorithmic}[1]
\STATE {\bfseries Input:} Required metrics $Y$, Actual metrics $Y_{act}$, Tolerance $tol$, Number of Performance $NP$

\STATE Initialize $Fitness = 1$
    \FOR{each metric $Y_i$ in $Y$}
        \IF{$Y_i \in Y_{A-class}$ (A-class metrics)}
            \STATE $Fitness -= \max\left(0, (Y_i - Y_{act,i})/{Y_i} - tol\right)/NP$
        \ELSIF{$Y_i \in Y_{B-class}$ (B-class metrics)}
            \STATE $Fitness -= \max\left(0, (Y_{act,i} - Y_i)/{Y_{act,i}}- tol\right)/NP$
        \ENDIF
    \ENDFOR
\STATE {\bfseries Output:} $Fitness$
\end{algorithmic}
\end{algorithm}
\begin{figure}[!t]
  \centering
  \includegraphics[width=\linewidth]{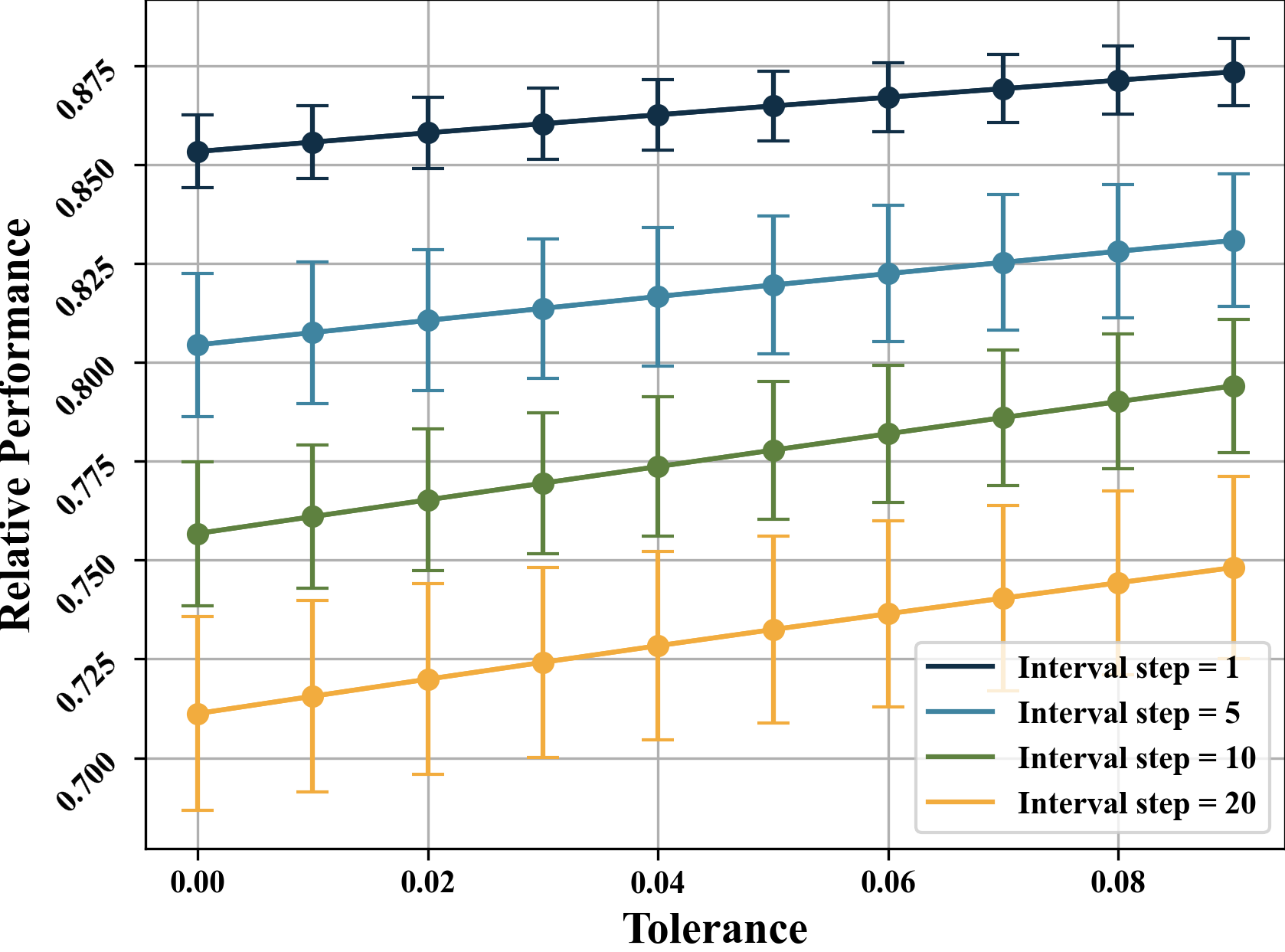}
  \caption{{\SolutionName}'s performance with different interval steps and tolerances.}
  \label{fig:Exp}
\end{figure}
\begin{table}[ht]
\centering
    \caption{{\SolutionName} performance under different settings}
    \label{tab:time}
    \begin{tabular}{c|cc}
        \toprule
        \textbf{\makecell{{\SolutionName}\\Setting}}&
        \textbf{\makecell{Relative\\Performance$\uparrow$}} & \textbf{\makecell{Time\\Consumption$\downarrow$}} \\ \hline
        \rowcolor{gray!20}Interval step@1  & $\mathbf{0.853\pm0.092}$ & $7.67s\pm0.15s$\\ 
        Interval step@5  & $0.804\pm0.181$ & $1.57s\pm0.08s$\\ 
        \rowcolor{gray!20}Interval step@10 & $0.757\pm0.182$ & $0.86s\pm0.08s$\\ 
        Interval step@20 & $0.711\pm0.244$ & $\mathbf{0.81s\pm0.09s}$\\ 
        \bottomrule
\end{tabular}
\end{table}
\subsection{Accuracy Evaluation}
Firstly, we will demonstrate the circuit generation accuracy of {\SolutionName} under different interval steps\cite{song2022denoisingdiffusionimplicitmodels}. Here, the interval steps are set to 1, 5, 10, and 20. For each interval step, we randomly sample 50 points across the \textbf{entire} sampling space, and then record their relative performance, which is defined in Algorithm.\ref{alg:Dynamic Loss}, and the time consumed. In Algorithm.\ref{alg:Dynamic Loss}, the required performance metrics are divided into two categories: Class-A and Class-B. Category A metrics, which are preferable when larger, include $Gain_{DC}$, $GBW$, $PM$, $SR_P$, $SR_N$, $VOH$, $CMRR$, and $PSRR$. Category B metrics, which are preferable when smaller, include $P_{diss}$, $VOL$, $Noise{@1kHz}$, and $Noise_{@1GHz}$. Worth to mention that, $C_L$ is set as the load when the amplifier is operating, which is used to examine various metrics of the circuit under this load. The average performance and standard deviation of the circuit generation of {\SolutionName} under these several interval steps are presented in Figure.\ref{fig:Exp}. (Note: For the sake of clarity in the drawing, we have uniformly divided the standard deviations by 10 here to prevent overlap and interference among them.) Furthermore, we have statistically examined the time expended by {\SolutionName} in generating a circuit under diverse interval steps. Incorporating its average circuit generation capacity and standard deviation at $tolerance = 0$, the outcomes are summarized in Table \ref{tab:time}.

\input{Experiment/expr_table.tex}

\input{Experiment/FOM_of_all_level}
\begin{figure}[t]
  \centering
  \includegraphics[width=\linewidth]{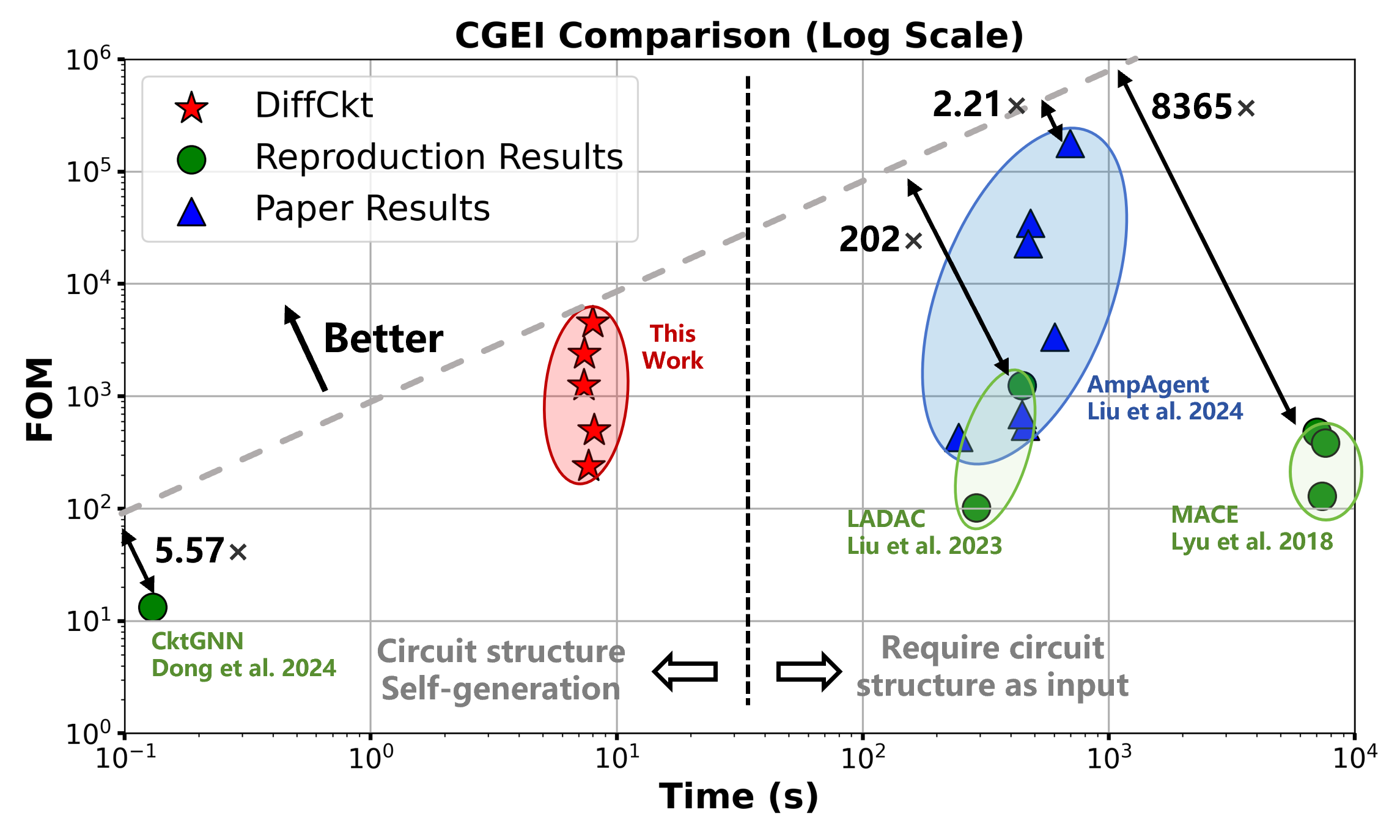}
  \caption{CGEI Comparison.}
  \label{fig:Exp2}
  \vspace{-15pt}
\end{figure}
To provide a more intuitive presentation, we have listed in Table \ref{tab: comp} the Actual Performance ($AP$) of three circuits generated by {\SolutionName} and the corresponding Required Performance ($RP$). This demonstrates that {\SolutionName} can successfully generate circuits with similar metrics based on the input $RP$.  

\subsection{Circuit Quality}
In this section, we will present the circuit generation quality of {\SolutionName} in different sampling spaces. The quality of the circuits is quantitatively evaluated using the Figure-of-Merit (FOM) value, which is widely used in the research of operational amplifiers\cite{peng2005transconductance,leung2001analysis,qu2016design}. It is defined as:
\[FOM = GBW (\text{MHz})\times C_L(\text{pF})/P_{diss}(\text{mW})\]
In addition, we also calculate the valid-generation ratio of {\SolutionName}, that is, to determine whether the operational amplifier has an amplification function. 
To observe these metrics of {\SolutionName}, we fixed the interval step at 1, randomly sampled 50 sets of data in each sampling space, and then tallied the FOM values, relative performance, and valid rates of the generated circuits. The statistical results are shown in Table.\ref{tab:FOM_of_all}. As can be seen from Table.\ref{tab:FOM_of_all}, when {\SolutionName} is given a set of higher required performance, it has the ability to generate circuits with higher FOM values. However, this leads to a slight decrease in the relative performance. When the input required performance is lower, {\SolutionName} exhibits higher relative performance and valid rate, indicating that the performance of the generated circuits can better meet the requirements. This is also consistent with our intuitive understanding. 

Furthermore, we conduct a quantitative comparison between {\SolutionName} and recent reproducible or comparable related works\cite{dong2024cktgnncircuitgraphneural,liu2024ladac,liu2024ampagent,pmlr-v80-lyu18a}. Here, we introduce the Circuit Generation Efficiency Index (CGEI), which is determined by both the figure of merit (FOM) of a single generated circuit and the time consumed. It is defined as:

\[CGEI=\frac{FOM (\text{MHz}\cdot\text{pF}/\text{mW})}{\text{Time consumption (s)}}\]

For \cite{dong2024cktgnncircuitgraphneural}, we trained it for 300 epochs on an A800 GPU. Since the original method used ideal modules, we referred to the process coefficients of TSMC65nm to convert the transconductance predicted by this work into power consumption, and then calculated its FOM value. Constrained by the VAE structure of this work, we input random matrices to the decoder part. Subsequently, we recorded the average FOM value of the finally generated circuit and the average time consumed.

For \cite{liu2024ladac}, we reproduced the work using GPT-4o, and conducted evaluations based on the two circuit structures and metrics mentioned in the original paper.

For \cite{pmlr-v80-lyu18a}, our experimental settings were configured as follows:
\begin{itemize}
\item \textbf{CPU}: Intel(R) Xeon(R) CPU E5 - 2698 v4 @ 2.20GHz
\item \textbf{Memory}: 440 GB
\item \textbf{Simulator}: Spectre(R)
\item \textbf{Optimization Settings}:
\begin{itemize}
\item \textbf{Maximum iterations}: 100 rounds
\item \textbf{Initial population size}: 40
\item \textbf{Batch size}: 40
\end{itemize}
\end{itemize}

Its input circuit structures consisted of the five topologies utilized in our data construction. We utilized three sets of metrics from Table.\ref{tab: comp} as input constraints for this method, with power consumption serving as the optimization metric. Additionally, we tallied the average generation quality and average time of the five circuits under these three sets of metrics for this method.

In addition, for work \cite{liu2024ampagent} that could not be replicated but were comparable, we converted their original IFOM values to corresponding FOM values (by dividing by the 1.8V supply voltage used in the original paper), and then tabulated the distribution of their metrics.

The comparison of various works is presented in Table \ref{tab:CGEI Comparison}. A visual comparison diagram can be found in Figure \ref{fig:Exp2}.
\begin{table}[t]
\centering
\caption{Detailed CGEI Comparison}
\label{tab:CGEI Comparison}
\resizebox{\linewidth}{!}{
\begin{tabular}{c|ccccc}
\toprule
Method&Index& FOM$\uparrow$ & \makecell{Time\\Consumption$\downarrow$}& CGEI$\uparrow$& \makecell{CGEI\\Comparison}\\
\hline
CktGNN \cite{dong2024cktgnncircuitgraphneural}&-& 13.27 & \textbf{0.13} &102&$5.57\times$\\
\hline
\multirow{2}{*}{LADAC \cite{liu2024ladac}}&1& 102 & 289 & 0.35&$1623\times$\\
&2& 1250 & 444& 2.81&$202\times$\\
\hline
&1& 478 & 7035 &0.068&$8365\times$\\
&2& 386 & 7612 &0.051&$11137\times$\\
\multirow{-3}{*}{MACE\cite{pmlr-v80-lyu18a}}&3& 129 & 7377 &0.017&$33411\times$\\ \hline
\multirow{2}{*}{AmpAgent\cite{liu2024ampagent}}&Worst& 458 & 545 & 1.19&$477\times$\\
&Best&179077 & 696 & 257.3&$2.21\times$\\ \hline
\rowcolor{green!10}&Worst& 86 & 7.69 & 11.2&$51\times$\\
\rowcolor{green!10}\multirow{-2}{*}{\textbf{DiffCkt}}&Best& \textbf{4530} & 7.97& \textbf{568}&-\\

\bottomrule
\end{tabular}
}

\end{table}

\text{\SolutionName} demonstrates superior performance even under evaluation conditions that are intentionally biased in favor of competing methods. Specifically:

\begin{itemize}
\item For \cite{pmlr-v80-lyu18a,liu2024ladac}: Requiring circuit structure as input (which \text{\SolutionName} autonomously generates)

\item For \cite{liu2024ampagent}: Exclude LLM execution time for Literature Analysis and Mathematics Reasoning (their critical path) + Using superior circuit structures from \text{\SolutionName}'s training dataset.
\item For \cite{dong2024cktgnncircuitgraphneural}: Only able to generate ideal behavior-level amplifier (vs. our transistor-level generation).

\end{itemize}
Even in this comparison that is disadvantageous to \(\text{\SolutionName}\), the best CGEI of {\SolutionName} has increased by $2.21 \sim 8365 \times$ compared to the best CGEI of related works. This fully attests to the high efficiency of \(\text{\SolutionName}\) in circuit generation and its attainment of the SOTA level.

%% file: Experiment/normal_tab.tex
\begin{table}[t]
\centering
\label{tab:Normal parameters_II}
\caption{Performance Metrics and Their Normalization Parameters}
\resizebox{\linewidth}{!}{
\begin{tabular}{l|c}
\toprule
Metric & Normalization Parameter \\ \midrule
\rowcolor{gray!20}Power Consumption($P_{diss}$) & $1\times10^{-3} W$ \\
DC Gain($Gain_{DC}$) & $100dB$ \\
\rowcolor{gray!20}Gain-Bandwidth Product($GBW$) & $10\times10^{6}Hz$ \\
Phase Margin($PM$) & $180$\textdegree \\
\rowcolor{gray!20}Positive Slew Rate($SR_P$)  & $10\times10^{6}V/s$ \\
Negative Slew Rate($SR_N$) & $10\times10^{6}V/s$ \\
\rowcolor{gray!20}Output Voltage Swing Low($VOL$) & $1.2V$ \\
Output Voltage Swing High($VOH$) & $1.2V$ \\
\rowcolor{gray!20}Common-Mode Rejection Ratio($CMRR$) & $100dB$ \\
Power Supply Rejection Ratio($PSRR$) & $100dB$ \\
\rowcolor{gray!20}Input Equivalent Noise @ 1 kHz($Noise_{@1kHz}$) & $1\times10^{-6}V/\sqrt{Hz}$ \\
Input Equivalent Noise @ 1 GHz($Noise_{@1GHz}$) & $1\times10^{-7}V/\sqrt{Hz}$ \\
\rowcolor{gray!20}Load Capacitance($C_L$) & $10\times10^{-12}F$ \\ \bottomrule
\end{tabular}}
\end{table}

%% file: Experiment/normal_table.tex
\begin{table}[!t]
\centering
\caption{Normalized Sampling Ranges for Circuit Metrics}
\label{tab:normalized}
\begin{tabular}{l|cccc}
\toprule
\textbf{Metric} & \textbf{External} & \textbf{High} & \textbf{Medium} & \textbf{Low} \\ \hline
\rowcolor{gray!20}\textbf{$P_{\text{diss}}$} & [0.03,0.20] & [0.05,0.35] & [0.35,0.65] & [0.65,1.00] \\
\textbf{$Gain_{DC}$} & [0.80,1.00] & [0.60,0.80] & [0.53,0.67] & [0.40,0.53] \\
\rowcolor{gray!20}\textbf{$GBW$} & [1.0,3.0] & [0.7,1.0] & [0.4,0.7] & [0.1,0.4] \\
\textbf{$PM$} & [0.31,0.33] & [0.31,0.33] & [0.28,0.31] & [0.25,0.28] \\
\rowcolor{gray!20}\textbf{$SR_P$} & [0.35,0.5] & [0.35,0.5] & [0.2,0.35] & [0.1,0.20] \\
\textbf{$SR_N$} & [0.35,0.5] & [0.35,0.5] & [0.2,0.35] & [0.1,0.2] \\
\rowcolor{gray!20}\textbf{$V_{OL}$} & [0.1,0.2] & [0.1,0.2] & [0.2,0.35] & [0.35,0.5] \\
\textbf{$V_{OH}$} & [0.8,0.9] & [0.8,0.9] & [0.65,0.8] & [0.5,0.65] \\
\rowcolor{gray!20}\textbf{$CMRR$} & [0.6,0.7] & [0.53,0.7] & [0.37,0.53] & [0.2,0.37] \\
\textbf{$PSRR$} & [0.6,0.7] & [0.53,0.7] & [0.37,0.53] & [0.2,0.37] \\
\rowcolor{gray!20}\textbf{$Noise_{@1kHz}$} & [0.5,0.67] & [0.5,0.67] & [0.67,0.83] & [0.83,1] \\
\textbf{$Noise_{@1GHz}$} & [0.5,0.67] & [0.5,0.67] & [0.67,0.83] & [0.83,1] \\
\rowcolor{gray!20}\textbf{$C_L$} & [1.0,2.0] & [0.7,1] & [0.4,0.7] & [0.1,0.4] \\
\bottomrule
\end{tabular}
\end{table}

%% file: Experiment/expr_table.tex
\begin{table}[t]
    \centering
    \caption{Generated Circuit Performance Comparison}
    \label{tab: comp}
    \resizebox{\linewidth}{!}{
    \begin{tabular}{c|cc|cc|cc}
    \toprule
    \multirow{2}{*}{\diagbox{Performance}{Test Num}}&\multicolumn{2}{c|}{Test Case 1}&\multicolumn{2}{c|}{Test Case 2}&\multicolumn{2}{c}{Test Case 3}\\
    \cline{2-7}
    &\makecell{RP}&\makecell{AP}&\makecell{RP}&\makecell{AP}&\makecell{RP}&\makecell{AP}\\
    \hline
    \rowcolor{gray!20}$C_L(pF)                   $&6.6&6.6&4.9&4.9&1.6&1.6\\
    $P_{diss}({\mu}W)\downarrow          $&158&\textbf{27}&74&\textbf{33}&68&\textbf{27}\\
    \rowcolor{gray!20}$Gain_{DC}(dB)\uparrow              $&65&\textbf{90}&45&\textbf{66}&45&\textbf{86}\\
    $GBW(MHz)\uparrow                  $&1.4&\textbf{2.3}&4.1&3.3&1.1&\textbf{1.5}\\
    \rowcolor{gray!20}$PM(\circ)\uparrow                 $&47&\textbf{53}&58&52&49&42\\
    $SR_P(V/{\mu}S)\uparrow            $&1.8&\textbf{3.3}&3.6&2.3&1.1&\textbf{1.13}\\
    \rowcolor{gray!20}$SR_N(V/{\mu}S)\uparrow            $&2.2&1.9&2.8&2.1&1.4&\textbf{1.58}\\
    $VOL(V)\downarrow                     $&0.18&\textbf{0.11}&0.32&\textbf{0.15}&0.28&\textbf{0.11}\\
    \rowcolor{gray!20}$VOH(V)\uparrow                    $&0.98&\textbf{1.1}&0.78&\textbf{0.83}&0.90&\textbf{1.0}\\
    $CMRR(dB)\uparrow                  $&56&55&50&\textbf{55}&53&\textbf{69}\\
    \rowcolor{gray!20}$PSRR(dB)\uparrow                  $&34&\textbf{68}&32&\textbf{34}&41&\textbf{59}\\
    $Noise_{@1kHz}({\mu}V/\sqrt{Hz})\downarrow $&0.54&\textbf{0.28}&0.6&\textbf{0.11}&0.85&\textbf{0.3}\\
    \rowcolor{gray!20}$Noise_{@1GHz}({\mu}V/\sqrt{Hz})\downarrow $&0.05&\textbf{0.01}&0.08&\textbf{0.03}&0.08&\textbf{0.02}\\
    \bottomrule
    \end{tabular}}
    
\end{table}

%% file: Experiment/FOM_of_all_level.tex
\begin{table}[!t]
\centering
\caption{Detailed {\SolutionName} under different sampling spaces}
\label{tab:FOM_of_all}
\resizebox{\linewidth}{!}{
\begin{tabular}{l|cccc}
\toprule
\textbf{Level} & \textbf{Best FOM$\uparrow$}& \textbf{FOM$\uparrow$} & \textbf{\makecell{Relative\\Performance$\uparrow$}} & \textbf{\makecell{Valid\\Rate$\uparrow$}} \\ \hline
\rowcolor{gray!20}External &\textbf{4530}& $\mathbf{2401 \pm 723}$ & $0.814 \pm 0.115$ & 88\% \\ 
High &1837& $1250 \pm 430$ & $0.845 \pm 0.119$ & 88\% \\ 
\rowcolor{gray!20}Medium &745& $499 \pm 207$ & $0.866 \pm 0.120$ & 84\% \\ 
Low &523& $238 \pm 153$ & $\mathbf{0.901 \pm 0.081}$ & \textbf{90\%} \\
\bottomrule
\end{tabular}}
\end{table}

%% file: Conclusion/conclusion.tex
In this paper, we present {\SolutionName}, a novel hybrid neural network system based on diffusion models for the automatic transistor-level generation of analog circuits. 
The dataset employed for model training, encompassing over 400k circuit samples, was sourced from TSMC 65nm CMOS technology. Notably, when the tolerance of the metric requirements is set at 0, the relative error of the generated circuit metrics can be, on average, constrained within $15\%$.
It is worth noting that, compared with other works, {\SolutionName} can significantly improve the circuit generation efficiency index (CGEI) by $2.21 \sim 8365 \times$, and reaches the SOTA level.

However, due to the inevitable randomness in sampling, {\SolutionName} currently cannot guarantee a 100-percent valid rate, which remains its limitation.

In the future, we will further transfer {\SolutionName} to other circuits and further enrich the diversity of the dataset to enable it to generate more diverse circuits. 

Through this paper, we aspire to pioneer a novel approach to the automated design of analog circuits using diffusion models. The latest amplifier dataset will be open source, and its preview version is available at {\Code}.